\documentclass[prd,aps,twocolumn,showpacs,preprintnumbers,amsmath,amssymb,nofootinbib,floats,floatfix,superscriptaddress]{revtex4-1}

\usepackage{amssymb,amsmath,bm,bbm}
\usepackage{graphicx}
\usepackage{dcolumn}
\usepackage[hyperfootnotes=false]{hyperref}
\usepackage{xspace}
\usepackage{xcolor}
\usepackage{url}
\usepackage[utf8]{inputenc}
\usepackage{epstopdf}
\usepackage[normalem]{ulem}

\allowdisplaybreaks

\newcommand{\bea}{\begin{eqnarray}}
\newcommand{\eea}{\end{eqnarray}}

\newcommand{\eq}[1]{Eq.~\eqref{#1}}

\newcommand{\be}{\begin{equation}}
\newcommand{\ee}{\end{equation}}
\newcommand{\hc}{\mbox{h.c.}}

\newcommand{\epsp}{\mbox{$\epsilon^\prime/\epsilon$}}

\begin{document}
\preprint{PSI-PR-19-18, UZ-TH 41/19, CERN-TH-2019-142, Nikhef/2019-041, INT-PUB-19-041}
\title{\boldmath Correlating $\epsilon^\prime/\epsilon$ to hadronic $B$ decays via $U(2)^3$ flavour symmetry}

\author{Andreas Crivellin}
\affiliation{Paul Scherrer Institut, CH--5232 Villigen PSI, Switzerland}
\affiliation{Physik-Institut, Universit\"at Z\"urich, Winterthurerstrasse 190, CH-8057 Z\"urich, Switzerland}

\author{Christian Gross}
\affiliation{Dipartimento di Fisica dell'Universit{\`a} di Pisa and INFN, Sezione di Pisa, Pisa, Italy}
\affiliation{Theoretical Physics Department, CERN, 1211 Geneve 23, Switzerland}

\author{Stefan Pokorski}
\affiliation{Institute of Theoretical Physics, Faculty of Physics, University of Warsaw, ul. Pasteura 5,
	PL-02-093 Warsaw, Poland}

\author{Leonardo Vernazza}
\affiliation{Nikhef, Science Park 105, NL-1098 XG Amsterdam, The Netherlands}

\begin{abstract}
There are strong similarities between charge-parity (CP) violating observables in hadronic $B$ decays (in particular $\Delta A^-_{\rm CP}$  in $B\to K\pi$) and direct CP violation in Kaon decays ($\epsilon^\prime$): All these observables are very sensitive to new physics (NP) which is at the same time CP and isospin violating (i.e. NP with complex couplings which are different for up quarks and down quarks). Intriguingly, both the measurements of $\epsilon^\prime$ and $\Delta A^-_{\rm CP}$ show deviations from their Standard Model predictions, calling for a common explanation (the latter is known as the $B\to K\pi$ puzzle). For addressing this point, we parametrize NP using a gauge invariant effective field theory approach combined with a global $U(2)^3$ flavor symmetry in the quark sector (also known as less-minimal flavour violation). We first determine the operators which can provide a common explanation of $\epsilon^\prime$ and $\Delta A^-_{\rm CP}$ and then perform a global fit of their Wilson coefficients to the data from hadronic $B$ decays. Here we also include e.g. the recently measured CP asymmetry in $B_s\to KK$ as well as the purely isospin violating decay $B_s\to\phi\rho^0$, finding a consistent NP pattern providing a very good fit to data. Furthermore, we can at the same time explain \epsp~for natural values of the free parameters within our $U(2)^3$ flavour approach, and this symmetry gives interesting predictions for hadronic decays involving $b\to d$ transitions.
\end{abstract}

\maketitle

\section{Introduction}
\label{intro}

Even though the Standard Model (SM) of particle physics has been tested to an astonishing precision within the last decades, it cannot be the ultimate theory describing the fundamental constituents and interactions of matter. For example, in order to generate the matter anti-matter asymmetry of the universe, the Sakharov criteria~\cite{Sakharov:1967dj} must be satisfied. One of these requirements is the presence of CP violation, which is found to be far too small within the SM~\cite{Cohen:1993nk,Gavela:1993ts,Huet:1994jb,Gavela:1994ds,Gavela:1994dt,Riotto:1999yt} whose only source of CP violation is the phase of the Cabibbo-Kobayashi-Maskawa (CKM) matrix. Therefore, physics beyond the SM with additional sources of CP violation is needed. 

Thus, CP violating observables are promising probes of new physics (NP) as they could test the origin of the matter anti-matter asymmetry of the universe. In this respect, direct CP violation in Kaon decays (\epsp) is especially relevant, as it is very suppressed in the SM, extremely sensitive to NP and can therefore test the multi TeV scale~\cite{Buras:2014zga}. Furthermore, recent theory calculations from lattice and dual QCD~\cite{Buras:2015xba,Buras:2015yba,Bai:2015nea,Kitahara:2016nld} show intriguing tensions between the SM prediction and the experimental measurement. In order to explain this tension,\footnote{Calculations using chiral perturbation theory~\cite{Cirigliano:2011ny,Pich:2004ee,Pallante:2001he,Gisbert:2017vvj,Gisbert:2018tuf} are  consistent with the experimental value but have large errors.} NP must not only violate CP but in general also isospin~\cite{Branco:1982wp} (i.e. couple differently to up quarks as to down quarks) in order to give a sizeable effect in \epsp~\cite{Aebischer:2018quc}.

Interestingly, there are also tensions between theory and data concerning CP violation in hadronic $B$ meson decays, including the long-standing $B\to K\pi$ puzzle~\cite{Gronau:1998ep,Buras:2003yc,Buras:2003dj,Buras:2004ub,Baek:2007yy,Fleischer:2007mq}. Recently, LHCb data~\cite{Aaij:2018tfw} increased this tension~\cite{Fleischer:2017vrb,Fleischer:2018bld}, and also the newly measured CP asymmetry in $B_s\to K^+K^-$~\cite{Aaij:2018tfw} points towards additional sources of CP violation, renewing the theoretical interest in these decays~\cite{Datta:2019tuj,Faisel:2018bvs}. Like for \epsp, both CP and isospin violation are in general required for solving this tension.
This can be achieved with NP in electroweak penguin operators~\cite{Fleischer:1995cg,Fleischer:2008wb,Baek:2009pa} that may for instance be generated in $Z'$ models~\cite{Barger:2009qs,Barger:2009eq}. Furthermore, the same NP effects can be tested in the theoretically clean purely isospin violating decays $B_s \to \phi \rho^0$ and $B_s \to \phi \pi^0$~\cite{Fleischer:1994rs,Buras:2003yc,Hofer:2010ee,Hofer:2012vc} where the former one has been measured recently~\cite{Aaij:2016qnm}, putting additional constraints on the parameter space.

These intrinsic similarities between \epsp~and hadronic $B$ decays suggest a common origin of the deviations from the SM predictions resulting in correlations among them. This can be studied in a model independent way within an effective field theory (EFT) approach. In order to connect $\epsilon^\prime/\epsilon$ ($s\to d$ transitions) to hadronic $B$ decays ($b\to s,d$ transitions) a flavour link
is obviously necessary. Here, we assume a global $U(2)^3$ flavor symmetry in the quark sector~\cite{Barbieri:1995uv,Barbieri:1997tu,Barbieri:2011fc,Barbieri:2011ci,Crivellin:2011fb,Barbieri:2012uh,Barbieri:2012bh}.\footnote{The $U(2)^3$ flavour symmetry is analogous to Minimal Flavour Violation~\cite{Chivukula:1987fw,Hall:1990ac,Buras:2000dm} (MFV) which uses a global $U(3)^3$ flavour symmetry instead~\cite{DAmbrosio:2002vsn}. However, $U(3)^3$ flavour is anyway strongly broken by the third generation Yukawa couplings to $U(2)^3$.}
As we will see, this flavour symmetry yields the desired flavour structure for the Wilson coefficients: it predicts a large phase (equal to the CKM phase) in Kaon decays, and the effect in $B$ physics only differs by a relative order one factor (if the corresponding CKM elements are factored out) but contains an additional free phase.

\section{Setup and Observables}

Here we discuss our setup and the predictions for the observables. The strategy for this is the following: We will start with \epsp~where we want to explaining the difference between experiment and the SM prediction. This will allow us to restrict ourselves to the limited set of operators which are capable of achieving this. We will then move to hadronic $B$ decays, pointing out the striking similarities with \epsp, and then establish our $U(2)^3$ flavour setup. 

The experimental value for direct CP violation in Kaon decays~\cite{Batley:2002gn,AlaviHarati:2002ye,Abouzaid:2010ny},
\begin{equation}
\left({\epsilon^\prime}/{\epsilon}\right)_{\rm exp}=(16.6\pm 2.3)\times 10^{-4}\,,
\end{equation}
lies significantly above the SM prediction from lattice QCD~\cite{Bai:2015nea,Buras:2015yba,Kitahara:2016nld} which is in the range $\left({\epsilon^\prime}/{\epsilon}\right)_{\rm SM}\simeq (1 - 2)\times 10^{-4} $, with an error of the order of $5 \times 10^{-4} $. Note that the lattice estimate is consistent with the estimated upper limit from dual QCD~\cite{Buras:2015xba}.

In the past years, many NP explanations of the ${\epsilon^\prime}/{\epsilon}$ discrepantly have been put forward (see e.g.~\cite{Buras:2015kwd,Buras:2016dxz,Bobeth:2017xry,Endo:2016tnu,Bobeth:2016llm,Blanke:2015wba,Buras:2015yca,Buras:2015jaq,Tanimoto:2016yfy,Kitahara:2016otd,Endo:2016aws,Crivellin:2017gks,Endo:2017ums,Chen:2018ytc,Chen:2018vog,Haba:2018byj,Haba:2018rzf,Matsuzaki:2018jui,Aebischer:2018csl,Chen:2018stt,Iguro:2019zlc}). 
Since here we want to perform an EFT analysis we consider the impact of the operators listed in Ref.~\cite{Aebischer:2018quc}. First of all, one sees that there are eight operators (plus their chirality flipped counter parts) which give numerically large effects in \epsp. 
We will focus on these operators in the following since, requiring an explanation of \epsp, the NP scale for the other operators must be so low that it would be in conflict with direct LHC searches. Furthermore~-- since we will consider a $U(2)^3$ setup~-- the Wilson coefficients of scalar and tensor operators contributing to Kaon physics are suppressed by the corresponding tiny Yukawa couplings of the first and second generation. Therefore, we are left with the Lagrangian
\begin{equation}
\begin{aligned}
{\cal L}_{\epsilon^\prime/\epsilon} = C_q^{VLR}O_q^{VLR}+\tilde C_q^{VLR}\tilde O_q^{VLR}+ L\leftrightarrow R
\end{aligned}
\end{equation}
with $q=u,d$ and the operators 
\begin{equation}
\begin{aligned}
O_q^{VLR}&=(\bar s_\alpha \gamma^\mu P_L d_\alpha) ( \bar q_\beta \gamma_\mu P_R q_\beta)\,,\\
\tilde O_q^{VLR}&=(\bar s_\alpha \gamma^\mu P_L d_\beta) ( \bar q_\beta \gamma_\mu P_R q_\alpha)\,,\\
\end{aligned}
\end{equation}
plus their chirality flipped counterparts. Here, $\alpha$ and $\beta$ are color indices and therefore $O_q^{VLR}$ ($\tilde O_q^{VLR}$) is a color singlet (triplet) operator. However, noting that one needs a violation of isospin (which is conserved in the left-handed quark current due to $SU(2)_L$ gauge invariance) we can omit the operators with flipped chiralities and the NP contribution to $\epsilon^\prime/\epsilon$ is approximately given by~\cite{Aebischer:2018quc,Aebischer:2018csl}
\begin{equation}
\begin{aligned}
\left(\dfrac{\epsilon^\prime}{\epsilon}\right)_{\rm NP}\approx& \ 1 \, {\rm TeV^2} \big(124 \ \textrm{Im} (C_{d}^{VLR}-C_{u}^{VLR})\\
&\qquad\left.+432 \ \textrm{Im} (\tilde C_{d}^{VLR}- \tilde C_{u}^{VLR})\right) \,.
\end{aligned}
\end{equation}
for a NP scale of 1 TeV.\footnote{Here we took again into account that for an enhanced effect NP should be isospin violating and neglected small isospin conserving contributions in the numerical factors.}

As outlined in the introduction, we want to study correlations between hadronic $B$ decays and \epsp~using a $U(2)^3$ flavour symmetry. In particular we want to address the $B\to K\pi$ puzzle. Here the experimental value for
\begin{equation}\label{deltaACPm}
\Delta A^-_{\rm CP} \equiv A_{\rm CP}(B^- \to \pi^0 K^-) - A_{\rm CP}(\bar B^0 \to \pi^+ K^-) \,,
\end{equation}
is~\cite{HFLAV16}
\begin{equation}
\Delta A^-_{\rm CP}|_{\rm exp}=(12.4 \pm 2.1) \%\,,
\end{equation}
which deviates from the SM prediction~\cite{Hofer:2010ee} 
\begin{equation}
\Delta A^-_{\rm CP}|_{\rm SM}=(1.8^{+4.1}_{-3.2})  \%\,,
\end{equation}
at the 2$\sigma$ level.\footnote{Ref.~\cite{Beaudry:2017gtw} performed a fit to all $B \to \pi K$ data and finds that the p-value crucially depends on the ratio of the color-suppressed to the color-allowed tree amplitudes. Since an acceptably good fit can be achieved if this ratio is somewhat larger than what is predicted from QCD factorization it is not absolutely clear that $B \to \pi K$ data points to NP, but it certainly leaves room for it. In the following we will investigate how NP can account for the measurement.} In addition, one has to take into account {also other CP asymmetries and total branching rations of hadronic $B$ decays involving $b\to s$ transitions. Here, } the experimental measurements of~\cite{Aaij:2018tfw,Aaij:2016qnm}
\begin{align} \label{otherobservables}
A^{}_{\rm CP}[B_s\to K^+K^-]_{\rm exp} &= (-20.0 \pm 6.0 \pm 2.0) \% \,,
\\*
{\rm Br}[B_s\to\phi\rho^0]_{\rm exp}&= (2.7 \pm 0.7 \pm 0.2 \pm 0.2) \times 10^{-7}\,,
\nonumber
\end{align}
which agree with the SM predictions
\begin{equation}
\begin{aligned}
A^{B_s}_{\rm CP}|_{\rm SM} &= (-5.9^{+26.6}_{-5.1})\% \,, \\
{\rm Br}[B_s\to\phi\rho^0]_{\rm SM}&= (5.3^{+1.8}_{-1.3}) \times 10^{-7}\,,
\end{aligned}
\end{equation}
 at the 1--2 $\sigma$ level, {are two of the most important examples in with respect to SM accuracy and experimental precision.}

For hadronic $B$ decays it is standard to use the effective Hamiltonian 
\be\label{effNP}
{\cal H}_{\rm eff}^{\rm NP} = -\frac{4G_F}{\sqrt{2}}
V_{tb}V^*_{ts} \!\!\!\! \sum_{q=u,d,s,c} 
\left( C^{q}_5 O^{q}_5 +  C^{q}_6 O^{q}_6 \right)+\hc\,,
\ee
for $b\to s$ transitions where the four-quark operators are defined as 
\be\label{a30c}
\begin{aligned}
	O_5^{q}&=(\bar{s}_\alpha \gamma^\mu P_L b_\alpha) \, (\bar{q}_\beta \gamma_\mu P_R q_\beta),\\
	O_6^{q}&=(\bar{s}_\alpha \gamma^\mu P_L b_\beta) \, (\bar{q}_\beta \gamma_\mu P_R q_\alpha) \,.
\end{aligned}
\ee
The corresponding expressions for $b\to d$ transitions follow by replacing $\bar s$ with $\bar d$ and $V_{tb}V^*_{ts}$ by $V_{tb}V^*_{td}$. Here, we consider only the operators motivated by \epsp,~as discussed in the last subsection, and neglect the numerically very small contributions of $q = c,s$ in~\eq{effNP}. 
Under the assumption of {a global} $U(2)^3$ flavour symmetry (to be discussed later on) the NP Wilson coefficients  carry a common new weak phase $\phi$ and we parameterise them as 
\be\label{WNPdef}
C^{d,u}_5 = c^{d,u}_5 \, e^{i \phi}\,, \quad \qquad 
C^{d,u}_6 = c^{d,u}_6 \, e^{i \phi}\,.
\ee

Like for \epsp, the leading effect which is necessary to account for the $K\pi$ puzzle is isospin violating. This can be easily seen by using an intuitive notation, similar to the one used in Ref.~\cite{Hofer:2010ee}. We parameterize the NP contribution to $K\pi$ decays in terms of $r_{\rm NP}^{q}$ ($r_{\rm NP}^{{\rm A},q}$), representing the ratio of NP penguin (annihilation) amplitudes with respect to the dominant QCD penguin amplitude of the SM. Therefore, one has for instance  
\begin{align} \label{DeltaACPparam}
\Delta A_{\textrm{CP}}^-&\simeq 
- 2 {\rm Im}(r_{\rm C}) \sin \gamma 
\\ 
&
\hspace{-1.2cm}
+ 2 \left[  
{\rm Im}(r_{\rm NP}^{d}) - {\rm Im}(r_{\rm NP}^{u})+ {\rm Im}(r_{\rm NP}^{{\rm A},d}) - {\rm Im}(r_{\rm NP}^{{\rm A},u}) 
 \right] \sin \phi, 
\nonumber
\end{align}
where $r_{\rm C}$ originating from the color suppressed tree topology amplitude of the SM. Here $\gamma$ is the CKM phase defined as $V_{ub} = |V_{ub}|e^{-i \gamma}$ and $\phi$ a generic weak phase of the NP contribution. We see that isospin violation is needed to get an effect in $\Delta A_{\textrm{CP}}^-$. Thus, interesting effects are expected in other {hadronic $B$ decays} sensitive to isospin violations, such as {the analogues of $\Delta A^-_{\rm CP}$ with $PV$ (pseudo-scalar and vector) and $VV$ (two vector) mesons in the final state (e.g. decays in which one replaces $\pi$ and $K$ in eq.~(\ref{deltaACPm}) with $\rho$ or $K^{*}$). Furthermore, an equivalent difference of direct CP asymmetries constructed for $B_s \to KK$ decays, i.e. $\Delta A_{\textrm{CP}}^{\rm KK} \equiv A_{\rm CP}(\bar B_s \to \bar K^0 K^0) - A_{\rm CP}(\bar B_s \to K^- K^+)$, and the purely isospin violating decays $B_s\to\phi\pi^0$ and $B_s\to\phi\rho^0$ are sensitive to isospin violating NP as well.} 

{The amplitudes of hadronic $B$ decays, like the ones involved in ratios $r_{\rm NP}^{q}$ and $r_{\rm NP}^{{\rm A},q}$ in \eq{DeltaACPparam} contain strong phases originating from QCD effects.} These phases can be calculated at next-to-leading order using QCD factorisation~\cite{Beneke:1999br,Beneke:2001ev,Beneke:2003zv}. This calculation is rather technical and {involves many input parameters} (see e.g. Refs.~\cite{Beneke:2009eb,Hofer:2010ee} for a detailed discussion on the calculation of NP operators matrix elements in the context of QCD factorisation). Thus we provide here semi-numerical formulas which describe the NP effect in the observables in appendix~\ref{appendix} {based on \eq{WNPdef} as input. However, these formula only serve as an illustration of impact of NP while in the phenomenological analysis we will perform a global fit (including also theory errors of the NP contributions), as done in Ref.~\cite{Hofer:2010ee}, to take all measurements consistently into account.}

{Let us now turn to the connection between \epsp \ and hadronic $B$ decays. For this we consider the $SU(2)_L$ invariant operators~\cite{Buchmuller:1985jz,Grzadkowski:2010es}}
\begin{align}
{\cal L}_{\rm SMEFT}&=\dfrac{1}{\Lambda^2}\left(C^{(1)ijkl}_{Qq}O^{(1)ijkl}_{Qq} +C^{(3)ijkl}_{Qq}O^{(3)ijkl}_{Qq} \right)
\end{align}
with 
\begin{align}
\begin{aligned}
O^{(1)ijkl}_{Qq}=\bar Q_i^\alpha\gamma^\mu P_L Q_j^\alpha \bar q_k^\beta\gamma_\mu P_R q_l^\beta\,,\\
O^{(3)ijkl}_{Qq}=\bar Q_i^\alpha\gamma^\mu P_L Q_j^\beta \bar q_k^\beta\gamma_\mu P_R q_l^\alpha\,,
\end{aligned}
\label{OSMEFT}
\end{align}
where $i,j,k,l$ are flavour indices, $q=u,d$ and $Q$ stands for the quark $SU(2)_L$ doublet. {Depending on the flavour structure, these operators enter \epsp~or hadronic $B$ decays.}

\begin{figure*}[t]
	\begin{center}
		\begin{tabular}{cp{7mm}c}
			\includegraphics[width=0.58\textwidth]{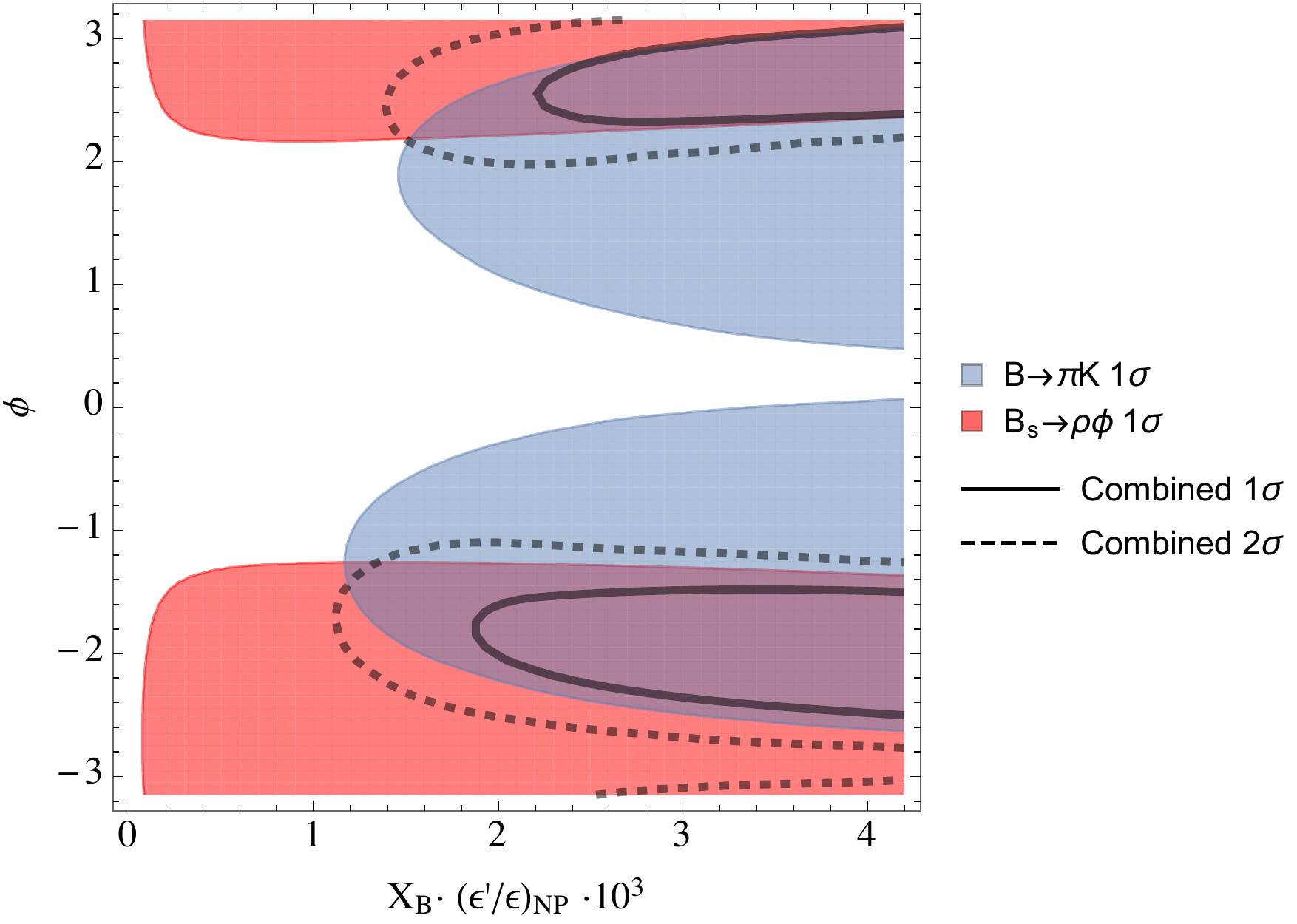}
			\includegraphics[width=0.413\textwidth]{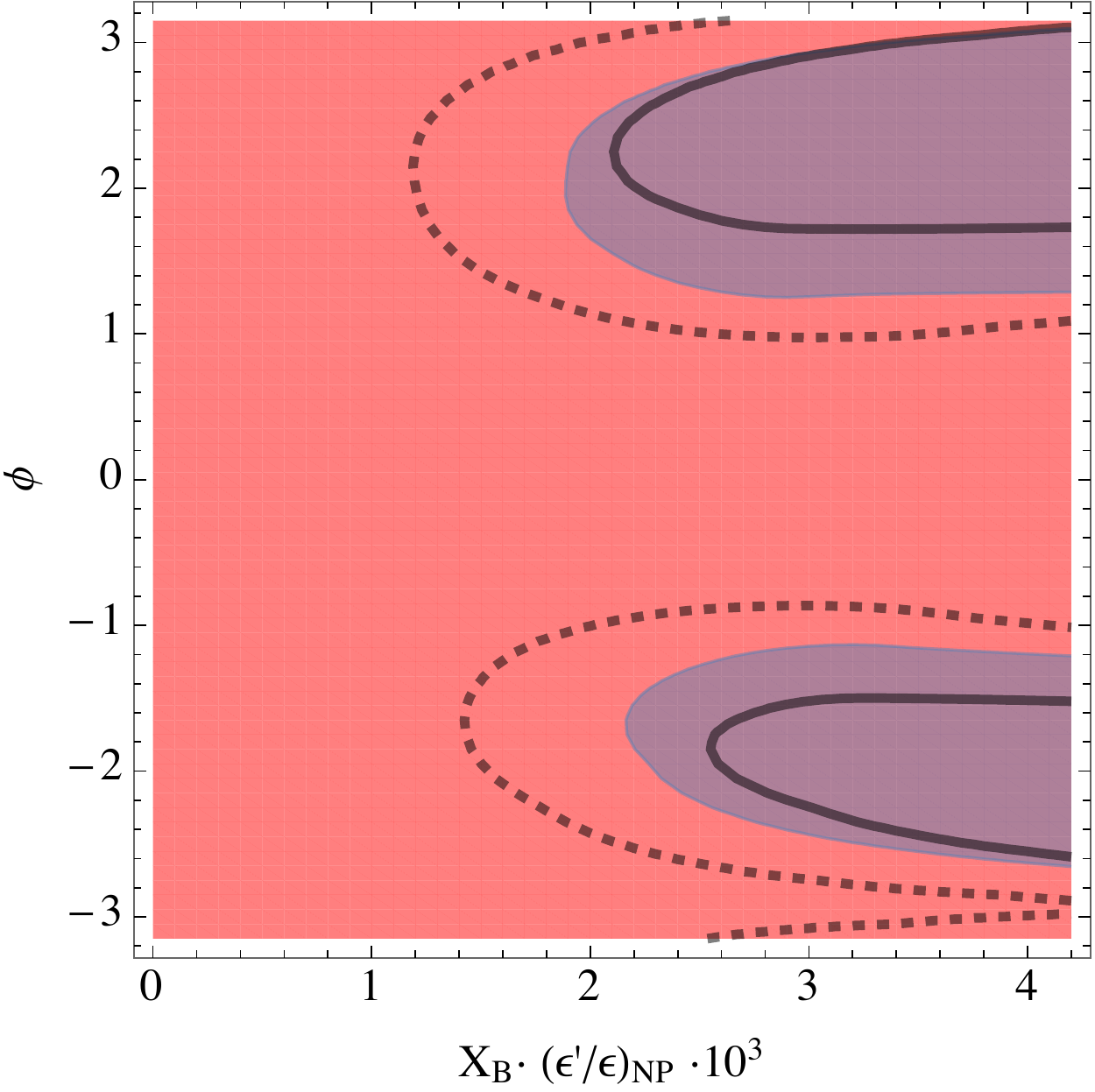}
		\end{tabular}
	\end{center}
	\caption{Preferred regions from hadronic $B$ decays in the $x_B\times (\epsilon^\prime/\epsilon)_{\rm NP}/ 10^{-3}$ vs. $\phi$ plane for the color singlet case (1) on the left and the color triplet case (3) on the right.  The preferred regions are obtained by marginalizing over $-0.12 <z^{\left(1\right)}<0.12$, and $-0.04 <z^{\left(3\right)}<0.04$. Note that all regions overlap at the $1\,\sigma$ level, resulting in a very good global fit (black bounded regions). Furthermore, one can explain the tensions in hadronic $B$ decays for an effect around $10^{-3}$ in $\epsilon^\prime/\epsilon$ (as suggested by the tension between SM  and experiment) for $x_B$ being of order one (as required by $U(2)^3$ flavour). \label{hadronicB}}         
\end{figure*}

Now, we employ the $U(2)^3$ flavour symmetry in the quark sector in order to link Wilson coefficients with different flavours to each other. First of all, note that with respect to the right-handed current we are only interested in the flavour diagonal couplings to $u,d$ and do not need to consider the couplings to heavier generations due to their suppressed effects in the observables. Concerning the left-handed current, $U(2)^3$ flavour with a minimal spurion sector predicts that { $s\to d$ transitions are proportional to $V_{ts}^*V_{td}$ while $b\to s(d)$ are proportional to $V_{ts(d)}^*V_{tb}$ and the relative effect is governed by an order one factor $x_B$ {and} a free phase $\phi$~\cite{Barbieri:2012uh}. Thus, \eq{OSMEFT} can be written as}
\begin{align}
\begin{aligned}
C_{Qq}^{(a)2111} &= V_{td}^{}V_{ts}^*c_q^{\left( a \right)}\\
C_{Qq}^{(a)2311} &= V_{tb}^{}V_{ts}^*{x_B}{e^{i\phi}}c_q^{\left( a \right)}\\
C_{Qq}^{(a)1311} &= V_{tb}^{}V_{td}^*{x_B}{e^{i\phi }}c_q^{\left( a \right)}
\end{aligned}
\end{align}
with $a=1,3$ (denoting the color singlet and triplet structure) and $q=u,d$. Note that due to the hermiticity of the operators in \eq{OSMEFT} $c_q^{\left( 1,3 \right)}$ must be real and that conventional MFV (based on $U(3)$ flavour) is obtained in the limit $\phi\to0$ and $ x_B\to 1$. Therefore, using MFV instead of $U(2)^3$ would provide an effect in \epsp~but {no source of CP violation in hadronic $B$ decays.}

With these conventions we obtain for the Wilson coefficients entering \epsp~and hadronic $B$ decays
\begin{align}
C_q^{VLR} &= \frac{{V_{ts}^*V_{td}c_q^{\left( 1 \right)}}}{{{\Lambda ^2}}}\,,
&
\!\!\!\!\!\!\tilde C_q^{VLR} &= \frac{{V_{ts}V_{td}^*c_q^{\left( 3 \right)}}}{{{\Lambda ^2}}}\,,
\\
C_5^q &= \frac{{\sqrt 2 }}{{4{G_F}{\Lambda ^2}}}{x_B}{e^{i\phi }}c_q^{\left( 1 \right)}\,,
&
\!\!\!\!\!\!C_6^q &= \frac{{\sqrt 2 }}{{4{G_F}{\Lambda ^2}}}{x_B}{e^{i\phi }}c_q^{\left( 3 \right)}\,. 
\nonumber
\end{align}

\section{Phenomenological Analysis}
\label{pheno}

{Here we present the results of the global fit to the data from hadronic $B$ decays. Taking into account that NP must have a common weak phase $\phi$ originating from $U(2)^3$ symmetry breaking we define}
\begin{equation}
{x^{\left( a \right)}} \equiv c_d^{\left( a \right)} - c_u^{\left( a \right)}\,,\qquad
{z^{\left( a \right)}} \equiv c_d^{\left( a \right)} + c_u^{\left( a \right)}\,,
\end{equation}
{for future convenience where ${x^{\left( a \right)}}$ (${z^{\left( a \right)}}$)  parametrizes the isospin violating (conserving) effects. Marginalizing over ${z^{\left( a \right)}}$ in the ranges from $-0.12 <z^{\left(1\right)}<0.12$, and $-0.04 <z^{\left(3\right)}<0.04$} we have three degrees of freedom for both the singlet scenario (1) and the triplet scenario (3). While the $\chi^2$ of the SM is 18.8, the best fit points for our two scenarios are
\begin{eqnarray} 
\begin{aligned}\label{fit-with-marginalisation}
x_B x^{(1)}=0.306\,, \quad  {x_B z^{\left( 1 \right)}} = -0.12 \,, \\
x_B x^{(3)}=0.144\,, \quad  {x_B z^{\left( 3 \right)}} = -0.04 \,, \\
\end{aligned}
\end{eqnarray}
with a phase of 
\begin{equation}
\phi^{(1)}=157.6^{\circ}\, , \qquad 
\phi^{(3)}=169.0^{\circ}\, ,
\end{equation}
and
\begin{equation}
\Delta \chi^2(1)=16.5\,, \quad  \Delta \chi^2(3)=13.7\,.
\end{equation}
This corresponds to pulls of $3.3\,\sigma$ for (1) and $2.9\,\sigma$ (3) with respect to the SM. Let us also consider the case in which $z^{\left( a \right)} = 0$, which corresponds to the scenario of maximal isospin violation. In this case the best fit points are $x_B x^{(1)}=0.312$, $\phi^{(1)}=163.3^{\circ}$, and $x_B x^{(3)}=0.142$, $\phi^{(3)}=-146.1^{\circ}$. The $\chi^2$ difference with respect to the SM are now $\Delta \chi^2(1)=15.3$ and $\Delta \chi^2(3)=12.9$, which corresponds to pulls of $3.5\,\sigma$ for (1) and $3.0\,\sigma$ for (3) with respect to the SM for two degrees of freedom. 

Now, we can correlate hadronic $B$ decays to \epsp. For this we observe that the NP contribution to \epsp~can be directly expressed in terms of $x^{(a)}$ as
\begin{align}
\left( {\frac{{\epsilon'}}{\epsilon}} \right)_{{\rm{NP}}} \!\!\!\!\!\! \approx \frac{0.018 \, x^{(1)}}{(\Lambda/\mathrm{TeV})^2} \,,
\quad 
\left( {\frac{{\epsilon'}}{\epsilon}} \right)_{{\rm{NP}}} \!\!\!\!\!\! \approx \frac{0.062 \, x^{(3)}}{(\Lambda/\mathrm{TeV})^2}\,,
\end{align}
for the color singlet and triplet case, respectively. Note that the phase of the contribution to \epsp~is fixed by the $U(2)^3$ flavour symmetry such that $\phi$ only enters in hadronic $B$ decays. Furthermore, $z^{(a)}$ is not correlated to \epsp~where only the difference $x^{(a)}$ enters and just a free parameter over which we will marginalize {as described above}. Therefore, we can express $x^{(a)}$ in terms of the NP contribution to \epsp~and {show the effects} in hadronic $B$ decays as a function of $x_B \times (\epsilon^\prime/\epsilon)_{\rm NP}/10^{-3}$ and $\phi$. 

\begin{figure*}[t]
	\begin{center}
		\begin{tabular}{cp{7mm}c}
			\includegraphics[width=0.6\textwidth]{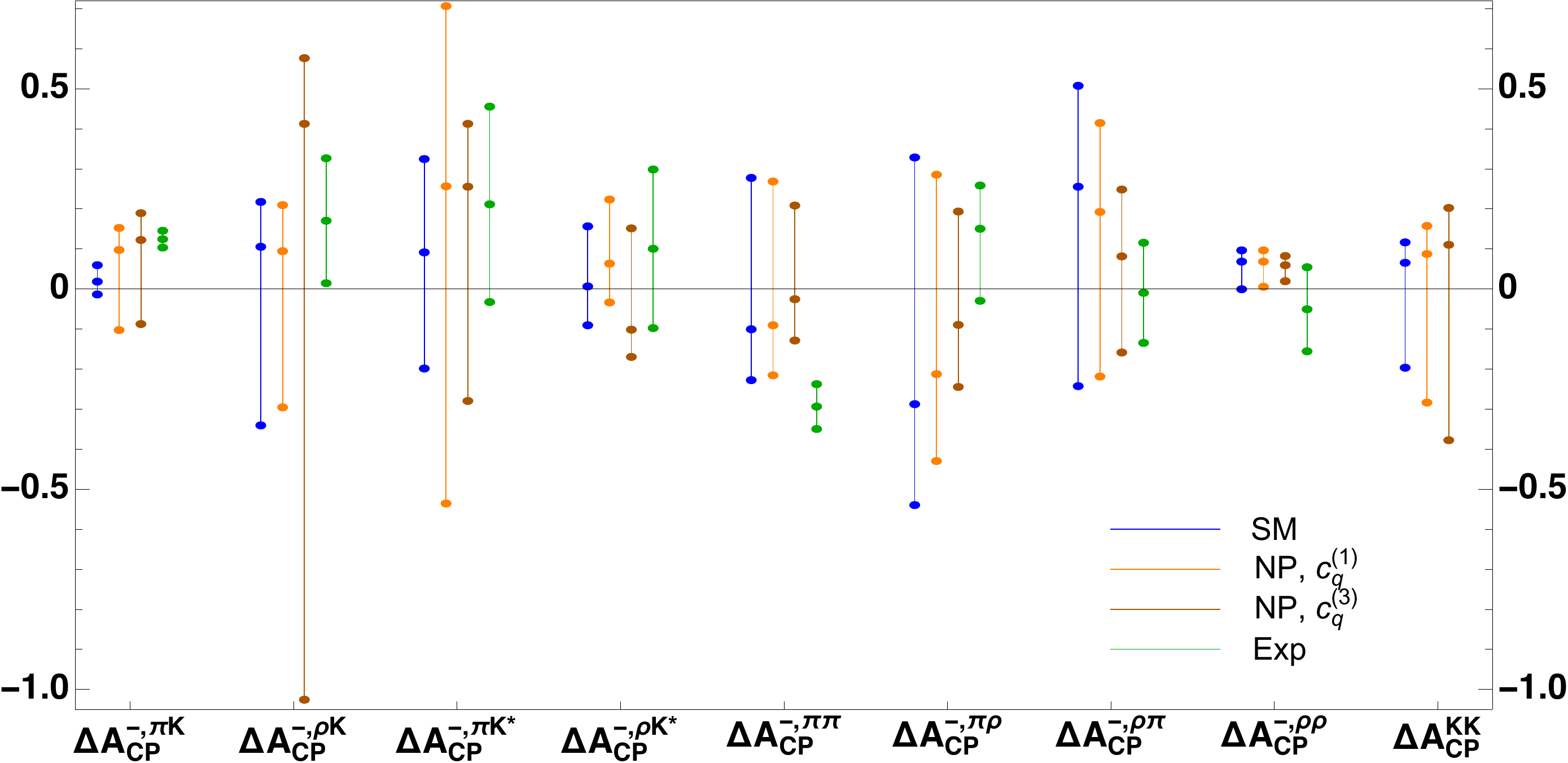}
		\end{tabular}
	\end{center}
	\caption{Predictions for differences of direct CP asymmetries. The first four observables are defined in Eq.~(\ref{ObsnpNum1b}) and involve $b \to s$ transitions while the following four are $b \to d$ transitions, predicted via the $U(2)^3$ flavour symmetry. The last observable $\Delta A^{KK}_{\rm CP}$ is defined in Eq.~(\ref{ObsnpNum1d}). It involves a $b \to s$ transition but no experimental result is available yet.  \label{hadronicBpredict}}         
\end{figure*}

The corresponding result is {depicted} in Fig.~\ref{hadronicB} where the preferred regions from hadronic $B$ decays are {displayed}. Note that all regions are consistent with each other (i.e. all overlap at the $1\,\sigma$ level), {such that} one can account for the deviations (mainly in $A^{}_{\rm CP}[B_s\to K^+K^-]_{\rm exp}$ and $\Delta A^-_{\rm CP}$) without violating bounds from other observables. 
From Fig.~\ref{hadronicB} one can also see that a natural order one value of $x_B$ can not only account the tensions in hadronic $B$ decays but also give a NP contribution to \epsp~of the order of $10^{-3}$ as needed to explain the tension.

In Fig.~\ref{hadronicBpredict} we show the predictions for various (differences of) CP asymmetries within the SM compared to the {one of the best fit points for the two scenarios as well as} and the corresponding experimental results. We use our $U(2)^3$ flavour symmetry to give predictions for hadronic $B$ decays involving $b\to d$ transitions as well. Although the fit clearly indicates isospin violating NP as the preferred solution to the $\Delta A_{\rm CP}$ problem, we notice that the errors of the theory predictions are still quite large, calling for future improvements in the calculational methods. Similarly, a clearer picture could be obtained with more precise experimental measurements, in particular for the $PV$
and $VV$ decay modes.

\section{Conclusions and Outlook}\label{conclusions}

{In this article we pointed out intrinsic analogies between \epsp~and CP violation in hadronic $B$ decays, in particular $\Delta A^-_{\rm CP}$: These observables are all sensitive to 4-quark operators with flavour changing neutral currents in the down sector and test the combined effects of CP and isospin violation. Therefore, the $B\to K\pi$ puzzle increases the interest in \epsp~and vice versa, calling for a combined explanation. 

After identifying the two operators which are capable of explaining the \epsp~anomaly within an $U(2)^3$ flavour setup we performed a global fit to the data from hadronic $B$ decays. We find that both operators provide a consistent pattern in hadronic $B$ decays resulting in a very good fit which is more than 3$\,\sigma$ better than the one of the SM. Furthermore, the $U(2)^3$ flavour symmetry is consistent with a common explanation of the anomalies in \epsp~and hadronic $B$ decays, providing at the same time interesting predictions for hadronic decays involving $b\to d$ transitions (such as $B\to K^+K^-$ and $B\to \pi\pi$) which can be tested experimentally in the near future by LHCb. However, further progress of the theory side is crucial in order to improve the precision of the theoretical results.}

\vspace{5pt}

{\it Acknowledgments} --- {\small 
We thank Andrzej Buras, Robert Fleischer and David Straub for useful discussions. We are grateful to Greg Landsberg for bringing the LHCb measurement of $B_s\to\phi\rho$ to our attention. The work of A.C. is supported by a Professorship Grant (PP00P2\_176884) of the Swiss National Science Foundation. C.G. is supported by the European Research Council grant NEO-NAT. S.P. research was supported by the Alexander von Humboldt Foundation. He thanks Slava Mukhanov for his hospitality at the LMU, Munich. L.V. is supported by the D-ITP consortium, a program of NWO funded by the Dutch Ministry of Education, Culture and Science (OCW). C.G. and L.V. thank the Mainz Institute for Theoretical Physics (MITP) of the DFG Cluster of Excellence PRISMA$^+$ (Project ID 39083149) for hospitality. A. C. thanks the INT at the University of Washington for its hospitality and the DOE for partial support during the completion of this work.
}

\appendix

\section{Additional non-leptonic decay observables}
\label{appendix}

In this appendix we collect semi-numerical formulae for other non-leptonic decay observables which are sensitive to isospin violating NP for the case of an $U(2)^3$ flavour symmetry. First of all, we list results for $\Delta A_{\textrm{CP}}^{-} $, and the corresponding observable obtained for $PV$ and $VV$ decays. One has  
\begin{widetext}
\begin{equation}
\label{ObsnpNum1b} 
\begin{aligned}
\Delta A_{\textrm{CP}}^{-,\pi K} &\simeq 0.02^{+0.04}_{-0.03}  
+ \big[ 13 (c_5^d - c_5^u) + 34 (c_6^d - c_6^u) \big] \sin \phi  
- \big[2 (c_5^d - c_5^u) + 5 (c_6^d  - c_6^u) \big] \cos \phi, 
\\
\Delta A_{\textrm{CP}}^{-,\rho K} &\simeq 0.11^{+0.11}_{-0.45}  
+ \big[21(c_5^d - c_5^u) + 39 (c_6^d - c_6^u) \big] \sin \phi  
- \big[ 12 (c_5^d - c_5^u) + 10 c_6^d - 1.1 c_6^u \big] \cos \phi,
\\
\Delta A_{\textrm{CP}}^{-,\pi K^*} &\simeq 0.09^{+0.23}_{-0.29}  
+ \big[ 23 (c_5^d - c_5^u) + 45 (c_6^d - c_6^u) \big] \sin \phi  
+ \big[-6 c_5^d + 8 c_5^u - 2 c_6^d  + 7 c_6^u \big] \cos \phi, 
\\
\Delta A_{\textrm{CP}}^{-,\rho K^*} &\simeq 0.01^{+0.15}_{-0.10}  
+ \big[ (c_5^d - c_5^u) - 20 c_6^d + 25 c_6^u \big] \sin \phi  
- \big[10 (c_5^d - c_5^u) + 2.5 c_6^d + 2.5 c_6^u \big] \cos \phi. 
\end{aligned}
\end{equation}
\end{widetext}
These formulae already include the evolution of the Wilson coefficients $C_{5,6}^u$ and $C^d_{5,6}$ in Eq.~(\ref{effNP}) from the electroweak scale to the scale $m_B$ and the numerical evaluation of the matrix elements using QCD factorization. Note also that the term $\propto \cos \phi$ in the direct CP asymmetries Eq.~(\ref{ObsnpNum1b}) originate from the interference between amplitudes proportional to $\gamma$ and $\phi$.

Next, we consider $B_s \to K K$ and related $VV$ decays. The CP observable in eq.~(\ref{otherobservables}) $A^{}_{\rm CP}[B_s\to K^+K^-]$ is given by
\begin{widetext}
\begin{equation}
\label{ObsnpNum1c} 
A_{\rm CP}[B_s\to K^+K^-] \simeq  -0.06^{+0.27}_{-0.05} 
+ \big[ -0.3 c_5^d + 2.6 c_5^u - 1.6 c_6^d  + 7.1 c_6^u \big] \sin \phi
+ \big[ - 0.75 c_5^u + 0.2 c_6^d - 2.3 c_6^u  \big] \cos \phi \,.
\end{equation}
More sensitive to isospin violation is the difference of direct CP asymmetries
\begin{equation}
\Delta A_{\textrm{CP}}^{\rm KK} \equiv A_{\rm CP}(\bar B_s \to \bar K^0 K^0) - 
A_{\rm CP}(\bar B_s \to K^- K^+)\,,
\end{equation}
and the equivalent difference defined for $VV$ modes. One has 
\begin{equation}
\label{ObsnpNum1d} 
\begin{aligned}
\Delta A_{\textrm{CP}}^{\rm KK} & \simeq 0.06^{+0.05}_{-0.26} 
+ \big[ 3 \big(c_5^d - c_5^u\big) + 9 \big(c_6^d  - c_6^u\big) \big] \sin \phi 
+\big[ c_5^u + 2 c_6^u \big] \cos \phi, 
\\
\Delta A_{\textrm{CP}}^{\rm K^*K^*} &\simeq -0.32^{+0.39}_{-0.05} 
+ \big[ (c_5^d - c_5^u) - 4 c_6^d + 3 c_6^u \big] \sin \phi 
+ \big[ 0.3 c_5^u - 0.5 c_6^d - 2.0 c_6^u \big] \cos \phi. 
\end{aligned}
\end{equation}
Last, we have the $B_s$ decays to $\pi,\phi$ and $\rho,\phi$, for which we have 
\begin{equation}
\label{ObsnpNum1e} 
\begin{aligned}
 {\rm Br}[B_s\to\phi\pi^0] &\simeq  \Big\{ 0.18^{+0.06}_{-0.05} 
 - \big[25 \big( c_5^d - c_5^u\big) + 8 \big(c_6^d - c_6^u \big) \big] \cos \phi 
 - \big[10 \big( c_5^d - c_5^u\big) + 2 \big( c_6^d - c_6^u \big) \big] \sin \phi \Big\} \times 10^{-6},  \\ 
{\rm Br}[B_s\to\phi\rho^0] &\simeq  
 \Big\{ 0.53^{+0.18}_{-0.13} \big[56 \big(c_5^d - c_5^u\big) + 18 \big( c_6^d - c_6^u\big) \big]   \cos \phi
+  \big[22 \big( c_5^d - c_5^u \big) + 6 \big( c_6^d - c_6^u\big) \big]  \sin \phi \Big\} \times 10^{-6}\,.
\end{aligned}
\end{equation}
\end{widetext}

\bibliography{BIB}

\end{document}